\documentclass[12pt]{article}

\title{A note about the quantum of area in a non-commutative space \\}
\author{Juan M. Romero, J.A. Santiago and J. David Vergara \\
\\
Instituto de Ciencias Nucleares, U.N.A.M.,\\
 Apdo. Postal 70-543, M\'exico D.F., M\'exico\\
\\
sanpedro, santiago, vergara@nuclecu.unam.mx }

\date{}

\begin{document}

\pagestyle{plain}

\maketitle

\begin{abstract}
In this note we show that in a two-dimensional non-commutative
space the area operator is quantized, this outcome is compared
with the result obtained by Loop Quantum Gravity methods.

\end{abstract}

One of the most relevant results of quantum mechanics is that
physical quantities as the energy and the angular momenta, can
have a discrete spectra. In fact, this theory begins with the
Planck idea about the quantum nature of light. In contrast,
general relativity modifies our understanding of gravity not as a
force but as geometry. In recent years, there are  several
attempts to reconcile the ideas of quantum mechanics and general
relativity. This implies in principle to change the geometrical
properties of the space-time at the Planck length scales
\cite{DeWitt}. One of this attempt is the loop quantum gravity
\cite{carlip:gnus}. A remarkable result of this theory, is that
geometrical operators as area or volume have discrete spectra at
the Planck length order. For instance, one can define an operator
for the area which has the spectrum

\begin{equation}
A (j_i)=8\pi \gamma l^{2}_{p} \sqrt{j_{i}\left( j_{i}+1\right) }
\label{aj1}
\end{equation}
where the $\left\{ j_{i}\right\} $ have integers and half-integers
values, $l_{p}=(\hbar G/c)^{1/2}=10^{-33} \hbox{cm}$ is the Planck
length  and $\gamma$ is a constant analogous to the Immirzi
parameter \cite{Merced}. The spectrum of the area operator is a
very important result of loop quantum gravity. However,
considering the recent numerical analysis of spin foam models
\cite{Baez}, it is possible that the correct area formula is given
not by equation (\ref{aj1}) but by
\begin{equation}
A\left( j_{i}\right) =\gamma l_{p}^{2}\left( j_{i}+1/2\right)   \label{aj2}
\end{equation}
This result was proposed in Ref. \cite{polychronakos:gnus} and our
analysis for a two dimensional non-commutative space agrees with
this spectra.

 On the other hand, the horizon area of black holes has been a
relevant geometrical object. The quantization of black holes was
proposed in the pioneering work of Bekenstein
\cite{bekenstein:gnus}(see the reviews \cite{bek1},\cite{bek2}). He
conjectured that the horizon area of a non extremal black hole
should have a discrete and uniformly spaced spectrum,
\begin{equation}
A_{n}=\tilde \gamma l_{p}^{2}n, \ \ \ \ \ \ \ \ \ n=1,2,...
\label{hod}
\end{equation}
where $\tilde \gamma $ is a dimensionless constant. Bekenstein
idea was to consider that the horizon area behaves as a classical
adiabatic invariant, and then use the Ehrenfest principle
\cite{ehrenfest:gnus}, which does correspond to any classical
adiabatic invariant a quantum operator with discrete spectrum.
Recently, Hod \cite{hod:gnus} gave arguments based in the Bohr's
Correspondence Principle, to determine the constant $\tilde \gamma
= 4 \ln 3$.

Another proposal to explain the physics of small scales is to
consider that the space is non commutative , {\it i. e.}, that the
commutation relations between spacial coordinates is of the form
\begin{equation}
\left[ \widehat{x}^{i},\widehat{x}^{j}\right] =i\hbar \Theta ^{ij}
\label{comm1}
\end{equation}
where $\Theta ^{ij}$ is an antisymmetric matrix with real and
constant parameters. The seminal idea of this proposal has been
attributed to Heisenberg \cite{jackiw:gnus}, although the first
report was given by Snyder \cite{Snyder:gnus}. Non-commutative
spaces appear in a natural way in the context of string theory
under some backgrounds \cite{witten:gnus}. Nevertheless, one can
also construct in a independent manner a field theory in a
non-commutative space \cite{szabo:gnus}. The physic associated to
this spaces is very interesting. For instance, in field the theory
build with these commutators exist a relationship between
ultraviolet and infrared divergences (mixture UV/IR)
\cite{minwalla:gnus}. In fact, this property could be required for
the formulation of a Field Theory valid at small and large scales
\cite{cohen:gnus}. In agreement with the bounds found for $\hbar
\Theta ^{ij}$, exist the possibility that the space may be
non-commutative before reaching the Planck scale
\cite{Carlson},\cite{irina:gnus}.

In this note we do a brief observation about the quantization of
the area operator by using non-commutative quantum theory in a two
dimensional Euclidean space. We begin with the commutation
relation
\begin{equation}
\left[ \widehat{x}_{1},\widehat{x}_{2}\right] =i\hbar \Theta  \ \
\ \ \ \ \ \ \ \Theta >0.  \label{comm2}
\end{equation}
Defining the operators

\begin{equation}
\widehat{a}=\frac{1}{\sqrt{2\hbar \Theta }}\left( \widehat{x}_{1}+i\widehat{x%
}_{2}\right) , \ \ \ \ \ \ \ \ \
\widehat{a}^{\dagger}=\frac{1}{\sqrt{%
2\hbar \Theta }}\left( \widehat{x}_{1}-i\widehat{x}_{2}\right).
\label{aa}
\end{equation}
These operators satisfy the following commutation relation

\begin{equation}
\left[ \widehat{a},\widehat{a}^{\dagger}\right] =1.  \label{comm3}
\end{equation}
Using $\widehat{a}$ and $\widehat{a}^{\dagger}$ we build the
number operator as

\begin{equation}
\widehat{N}=\widehat{a}^{\dagger}\widehat{a,}  \label{num}
\end{equation}
and thus we find
\begin{equation}
\left[ \widehat{N},\widehat{a}^{\dagger}\right]
=\widehat{a}^{\dagger} , \ \ \ \ \ \ \left[
\widehat{N},\widehat{a}\right] =-\widehat{a} \ \ . \label{num2}
\end{equation}
In this way, \ if $\ \widehat{N}|n\rangle =n|n\rangle $,  we
identify to $\widehat a^{\dagger}$ and $\widehat{a}$ as creation
and annihilation operators respectively.

 In terms of the coordinate
operators we can write
\begin{equation}
\widehat{N}=\frac{1}{2\hbar \Theta }\left( \widehat{x}_{1}^{2}+\widehat{x}%
_{2}^{2}-\hbar \Theta \right) ,  \label{N}
\end{equation}
therefore, if we define the operator
$$\frac{\widehat A}{\pi} \equiv
\widehat{x}_{1}^{2}+\widehat{x}_{2}^{2},$$ it now follows from
Eq.( \ref{N}) that
\begin{equation}
\frac{\widehat A}{\pi}=2\hbar \Theta \left( \widehat{N}+1/2\right)
,  \label{d2}
\end{equation}
Taking into account that the operator $\widehat A/\pi$ is positive
by definition, the operator  $\widehat{N}$ has a minimal
eigenvalue. In consequence  $\widehat A/\pi$ is  quantized  with
levels equally spaced by the interval $2\hbar \Theta$,
\begin{equation}
\frac{\widehat A}{\pi}| n\rangle
=\frac{A_{n}}{\pi}|n\rangle=2\hbar \Theta \left( n+1/2\right)
|n\rangle , \ \ n=0,1,... \label{dspec}
\end{equation}
with
\begin{equation}
|n\rangle = \frac{(\widehat a^\dagger )^n}{\sqrt{n!}} |0 \rangle
\end{equation}
It is tempting consider to $\sqrt{\widehat A /\pi}$ as a distance
operator. However, the non-commutativity of the coordinates in
this space implies that the definition of the notion of points and
the distance between them is no precise. Thus, is more suitable to
relate the operator with the area of a circumference. From the
definition of area of a circumference
$$A=\pi [(x_{1})^{2}+( x_{2})^{2}],$$
the quantum operator associated with it, is given by
$$\widehat A=\pi [(\widehat x_{1})^{2}+(\widehat x_{2})^{2}]$$
Thus, we conclude that the area has a discrete spectra in terms of
$2\pi\hbar\Theta$:
\begin{equation}
A_{n}=2\pi\hbar \Theta \left(
n+1/2\right)
\end{equation}
Starting in a completely different way, we have obtained the
spectra that looks like the black hole result Eq.$\left(
\ref{hod}\right) $ and also contains in part, the loop quantum
gravity result Eq.$\left( \ref{aj2}\right)$. In the original model
of Snyder \cite{Snyder:gnus}, also appears the quantization of
geometric quantities. It is interesting to observe that the
hypothesis of non commutativity of the space implies directly a
discrete spectrum for the area operator, something similar occurs
in the case of the distance where using the approach of Connes of
noncommutative geometry was obtained a discrete spectrum for the
distance operator \cite{Rovelli}.

It is remarkable that using the simple hypothesis of no
commutativity in two spatial dimension we can get a spectrum very
similar to the obtained by the proposal of loop quantum gravity
given in \cite{polychronakos:gnus} and also that this spectrum is
similar to the obtained for a black hole. This could be suggest
that the proposal given in \cite{polychronakos:gnus} is the
correct and that exist a very profound relationship between
quantum gravity and non-commutative geometry.

\end{document}